\documentclass[12pt, a4paper]{article}
\usepackage{graphicx}
\usepackage{amssymb}
\usepackage{amsmath}
\usepackage{bm}
\usepackage{color}
\usepackage{theorem}
\usepackage{subcaption}
\usepackage{listings}
\usepackage{colortbl}
\usepackage{tabularx}
\usepackage{longtable}
\usepackage[utf8]{inputenc}
\usepackage[T1]{fontenc}
\usepackage{lmodern}
\usepackage[top=30truemm,bottom=30truemm,left=25truemm,right=25truemm]{geometry}
\usepackage{comment}

\usepackage[sort&compress,numbers, merge]{natbib}

\definecolor{Orange}{cmyk}{0,0.61,0.87,0}
\definecolor{JungleGreen}{cmyk}{0.99,0,0.52,0}
\definecolor{OliveGreen}{cmyk}{0.64,0,0.95,0.40}
\definecolor{Brown}{cmyk}{0,0.81,1,0.60}
\definecolor{RoyalBlue}{cmyk}{0.71,0.53,0,0.12}
\definecolor{Gray}{cmyk}{0,0,0,0.40}
\definecolor{LightPink}{cmyk}{0.0,0.25,0,0}
\definecolor{LLightPink}{cmyk}{0.0,0.10,0,0}
\definecolor{LightBlue}{cmyk}{0.25,0,0,0}
\definecolor{LightGray}{cmyk}{0,0,0,0.2}

\allowdisplaybreaks[1]

\usepackage[colorlinks=true, linkcolor=OliveGreen, citecolor=RoyalBlue,
urlcolor=RoyalBlue]{hyperref}

\renewcommand{\thefootnote}{\fnsymbol{footnote}}

\begin{document}

\begin{titlepage}

  \begin{flushright}

\end{flushright}

\vskip 1.35cm
\begin{center}

{\Large
{\bf
From the LZ Event to Grand Unification:\\[5pt] 
Heavy Higgsinos and Proton Decay
}
}

\vskip 1.5cm

Shihwen Hor$^{a,b}$\footnote{
  E-mail address: \href{mailto:shihwen@sjtu.edu.cn}{\tt shihwen@sjtu.edu.cn}}, 
Natsumi Nagata$^c$\footnote{
E-mail address: \href{mailto:natsumi@hep-th.phys.s.u-tokyo.ac.jp}{\tt natsumi@hep-th.phys.s.u-tokyo.ac.jp}},
Tsutomu T. Yanagida$^{d}$\footnote{
  E-mail address: \href{mailto:tsutomu.tyanagida@sjtu.edu.cn}{\tt tsutomu.tyanagida@sjtu.edu.cn}}

\vskip 0.8cm

{\it ${}^a$Tsung-Dao Lee Institute, Shanghai Jiao Tong University,\\ 
No. 1 Lisuo Road, Pudong New Area, Shanghai, 201210, China} \\[3pt]
{\it ${}^b$School of Physics and Astronomy, Shanghai Jiao Tong University,\\
800 Dongchuan Road, Shanghai, 200240, China} \\[3pt]
{\it ${}^c$Department of Physics, University of Tokyo, Bunkyo-ku, Tokyo 113--0033, Japan} \\[3pt]
{\it ${}^d$Kavli Institute for the Physics and Mathematics of the Universe (Kavli IPMU), University of Tokyo, Kashiwa 277--8583, Japan}

\date{\today}

\vskip 1.cm

\begin{abstract}

 The recently reported LUX-ZEPLIN (LZ) event has renewed interest in inelastic Higgsino dark matter. While a weak-scale Higgsino provides a simple interpretation of the event, such a scenario is in tension with the IceCube bound from dark-matter capture and annihilation in the Sun. This tension can be avoided for a much heavier Higgsino, with a mass of order \(10^5\,\mathrm{GeV}\) and a neutral-state mass splitting of \(\simeq 400~\mathrm{keV}\), which in turn points to gaugino masses around \(10^7\,\mathrm{GeV}\). Such a heavy supersymmetric spectrum is often thought to be unfavorable for supersymmetric grand unified theories (GUTs), since it tends to worsen gauge coupling unification and lower the unification scale, potentially leading to excessively rapid proton decay. In this work, we revisit this expectation in minimal supersymmetric SU(5), taking into account the GUT-scale threshold corrections. We show that the resulting unification conditions allow the SU(5) gauge bosons to remain sufficiently heavy when the adjoint-Higgs self-coupling is small, thereby suppressing dimension-six proton decay. At the same time, the color-triplet Higgs mass can remain near the conventional GUT scale, \(M_{H_C}\sim 10^{16}\,\mathrm{GeV}\), particularly when the gluino is significantly heavier than the wino. Remarkably, in this region the dimension-five decay mode \(p\to K^+\bar\nu\) can have a lifetime within the reach of next-generation proton-decay searches, including Hyper-Kamiokande, JUNO, and DUNE. We also discuss supersymmetry-breaking scenarios that can give rise to the required supersymmetric mass spectrum.
  
\end{abstract}

\end{center}
\end{titlepage}

\renewcommand{\thefootnote}{\arabic{footnote}}
\setcounter{footnote}{0}

\section{Introduction}
\label{sec:introduction}

The recently reported LUX-ZEPLIN (LZ) event at a recoil energy of \(\simeq 248~{\rm keV}\)~\cite{LZ:2026axp} has motivated renewed interest in inelastic dark-matter scattering. A particularly simple interpretation is provided by Higgsino dark matter, for which the two neutral components are nearly degenerate and couple off-diagonally to the \(Z\) boson. A mass splitting of a few hundred keV can naturally shift the recoil spectrum toward the observed energy range, making the Higgsino an attractive candidate for explaining the event~\cite{Fan:2026kxx, Freese:2026sga, Wu:2026nhi}. For supersymmetric realizations of this interpretation and related complementary phenomenology, see Refs.~\cite{Yin:2026jnn, Du:2026guj, Bisal:2026khf, Cheung:2026byg, Langhoff:2026ujr, Frolovsky:2026tvq, Khan:2026zuj, Unwin:2026qep, Roy:2026wzb}.

This interpretation, however, is subject to a strong constraint from dark-matter capture in the Sun~\cite{Pospelov:2026ewn}. Higgsinos captured via inelastic scattering can subsequently annihilate and produce energetic neutrinos, leading to stringent bounds from IceCube~\cite{IceCube:2025fcu}. Subsequent independent analyses have reached similar conclusions, obtaining lower bounds on the Higgsino mass splitting of \(\simeq 0.5\)--\(0.6~{\rm MeV}\)~\cite{DiMauro:2026dqp,Bose:2026ndd,Nguyen:2026lui}. This tension can nevertheless be avoided if the Higgsino is much heavier than in the conventional thermal scenario.\footnote{For an alternative way to evade the IceCube limit based on an asymmetric dark matter scenario, see Ref.~\cite{Nagata:2026pbj}.} In particular, for the Higgsino mass of \(\simeq 10^5~{\rm GeV}\) and the Higgsino mass splitting of \(\simeq 400~{\rm keV}\), the LZ event can be accommodated while evading the solar-neutrino constraint~\cite{Langhoff:2026ujr, Luengas:2026tlf}.

Such a small splitting for a very heavy Higgsino has an immediate implication for the supersymmetric spectrum. In the minimal supersymmetric Standard Model (MSSM), the splitting between the two neutral Higgsino states is generated by mixing with the bino and wino and is approximately suppressed by their masses~\cite{Nagata:2014wma}. As it turns out, the splitting relevant for the LZ event points to electroweak gaugino masses of \(M_{1,2} = \mathcal{O} (10^7)~{\rm GeV}\). This raises an interesting question for supersymmetric grand unified theories (GUTs)~\cite{Dimopoulos:1981zb, Sakai:1981gr}.\footnote{The implications of the LZ event for supersymmetric GUTs have also been discussed in Ref.~\cite{Yin:2026jnn}.} Very heavy Higgsinos and gauginos are often regarded as unfavorable for supersymmetric grand unification, since they modify the running of the gauge couplings and may lower the GUT scale. The latter, in particular, could enhance proton decay mediated by the heavy GUT gauge bosons beyond the experimental limits.

In this paper, we show that this expectation need not hold even in minimal supersymmetric SU(5)~\cite{Dimopoulos:1981zb,Sakai:1981gr}. As discussed in Refs.~\cite{Hisano:1992mh,Hisano:1992jj}, gauge coupling unification, including threshold corrections at the supersymmetry and GUT scales, constrains two GUT-scale mass parameters: the color-triplet Higgs mass, \(M_{H_C}\), and \(M_{\mathrm{GUT}} \equiv (M_X^2 M_\Sigma)^{1/3}\), where \(M_X\) and \(M_\Sigma\) denote the masses of the SU(5) gauge bosons and the adjoint Higgs fields, respectively. Evolving the measured gauge couplings to high energies therefore allows these two mass parameters to be determined. Importantly, however, this does not uniquely determine the SU(5) gauge-boson mass, since only the combination \(M_{\mathrm{GUT}}\) is fixed. Consequently, if the self-coupling of the adjoint Higgs is sufficiently small, the adjoint Higgs can be much lighter than the gauge bosons, allowing \(M_X\) to remain sufficiently large. The resulting dimension-six proton decay rate can then readily satisfy the current experimental bound despite the heavy Higgsino and gaugino spectrum.

The color-triplet Higgs mass, on the other hand, is separately determined by the unification condition. We find that it can remain around the conventional supersymmetric GUT scale, \(M_{H_C}\sim 10^{16}~{\rm GeV}\), when the gluino is significantly heavier than the wino~\cite{Hisano:2013cqa}. In this region, the dimension-five~\cite{Sakai:1981pk, Weinberg:1981wj} decay mode \(p\to K^+\bar{\nu}\) can have a lifetime consistent with the current bound while remaining within the prospective reach of next-generation proton-decay searches~\cite{Dev:2022jbf}, including Hyper-Kamiokande~\cite{Hyper-Kamiokande:2018ofw}, JUNO~\cite{JUNO:2021vlw}, and DUNE~\cite{DUNE:2020fgq}. Thus, rather than being incompatible with supersymmetric grand unification, the heavy spectrum suggested by the LZ event can give rise to experimentally testable proton-decay signatures.

We also discuss supersymmetry-breaking scenarios that can realize the required supersymmetric mass spectrum. Our results provide an intriguing link between a possible dark-matter signal at the hundred-keV recoil scale, physics near the grand-unification scale, and more fundamental physics underlying supersymmetry and its breaking.

The rest of this paper is organized as follows. In Sec.~\ref{sec:mssu5}, we briefly review minimal supersymmetric \(\mathrm{SU}(5)\) and discuss how the color-triplet Higgs mass, \(M_{H_C}\), and the effective GUT scale, \(M_{\rm GUT}\), are determined from gauge coupling unification including threshold corrections. In Sec.~\ref{sec:protondec}, we study proton decay in the heavy-Higgsino scenario. In Sec.~\ref{sec:scenarios}, we discuss the theoretical implications of the heavy-Higgsino scenario, including supersymmetry-breaking mechanisms that can realize the required mass spectrum as well as cosmological scenarios that yield the correct Higgsino dark-matter abundance.\footnote{It was pointed out in Ref.~\cite{Feldstein:2013uha} that such heavy Higgsino dark matter could be detectable in xenon-based direct-detection experiments.} Finally, Sec.~\ref{sec:conclusion} is devoted to our conclusions.

\section{Minimal supersymmetric SU(5)}
\label{sec:mssu5}

In this section, we briefly review the minimal supersymmetric \(\mathrm{SU}(5)\) GUT~\cite{Dimopoulos:1981zb,Sakai:1981gr} and fix our notation and conventions. We then discuss the constraints on the GUT-scale mass spectrum obtained from gauge coupling unification, following Refs.~\cite{Hisano:1992mh,Hisano:1992jj,Hisano:2013cqa}.

\subsection{Minimal supersymmetric \texorpdfstring{\(\mathrm{SU}(5)\)}{SU(5)}}

The MSSM matter superfields are embedded into three generations of \(\overline{\mathbf{5}}\oplus\mathbf{10}\) representations of \(\mathrm{SU}(5)\). We denote the corresponding chiral superfields by \(\Phi_i\) and \(\Psi_i\), respectively, where \(i=1,2,3\) is the generation index. Their components are
\begin{align}
\Phi_i &=
\begin{pmatrix}
\bar D_{i1}\\
\bar D_{i2}\\
\bar D_{i3}\\
E_i\\
-N_i
\end{pmatrix}, \qquad 
\Psi_i
=
\frac{1}{\sqrt{2}}
\begin{pmatrix}
0 & \bar U_{i3} & -\bar U_{i2} & U_i^1 & D_i^1\\
-\bar U_{i3} & 0 & \bar U_{i1} & U_i^2 & D_i^2\\
\bar U_{i2} & -\bar U_{i1} & 0 & U_i^3 & D_i^3\\
-U_i^1 & -U_i^2 & -U_i^3 & 0 & \bar E_i\\
-D_i^1 & -D_i^2 & -D_i^3 & -\bar E_i & 0
\end{pmatrix},
\end{align}
where
\begin{equation}
L_i=
\begin{pmatrix}
N_i\\ E_i
\end{pmatrix},
\qquad
Q_i^a=
\begin{pmatrix}
U_i^a\\ D_i^a
\end{pmatrix},
\end{equation}
and \(a=1,2,3\) denotes the color index.

The two MSSM Higgs doublets are embedded into a pair of \(\mathbf{5}\) and \(\overline{\mathbf{5}}\) chiral superfields,
\begin{equation}
H=
\begin{pmatrix}
H_C^1\\
H_C^2\\
H_C^3\\
H_u^+\\
H_u^0
\end{pmatrix},
\qquad
\bar H=
\begin{pmatrix}
\bar H_{C1}\\
\bar H_{C2}\\
\bar H_{C3}\\
H_d^-\\
-H_d^0
\end{pmatrix},
\end{equation}
where
\begin{equation}
H_u=
\begin{pmatrix}
H_u^+\\ H_u^0
\end{pmatrix},
\qquad
H_d=
\begin{pmatrix}
H_d^0\\ H_d^-
\end{pmatrix}.
\end{equation}
The fields \(H_C^a\) and \(\bar H_{Ca}\) are the color-triplet Higgs superfields.

The gauge sector consists of an adjoint vector superfield \({\cal V}^A\), \(A=1,\ldots,24\), which contains the MSSM gauge multiplets as well as the \(X\) and \(Y\) gauge bosons associated with the broken generators of \(\mathrm{SU}(5)\). We collectively denote the latter by \(X\) bosons,
\begin{equation}
(X)_a^\alpha
\equiv
\begin{pmatrix}
X_a\\
Y_a
\end{pmatrix},
\end{equation}
where \(\alpha=1,2\) denotes the \(\mathrm{SU}(2)_L\) index.

The breaking of \(\mathrm{SU}(5)\) to the Standard Model (SM) gauge group, \(\mathrm{SU}(5) \to  \mathrm{SU}(3)_C\otimes\mathrm{SU}(2)_L\otimes\mathrm{U}(1)_Y\), is triggered by an adjoint Higgs chiral superfield \(\Sigma=\Sigma^A T^A\). Under the SM gauge group, its physical components include an \(\mathrm{SU}(3)_C\) octet \(\Sigma_8\), an \(\mathrm{SU}(2)_L\) triplet \(\Sigma_3\), and a singlet \(\Sigma_{24}\), in addition to the components eaten by the massive \(X\) gauge multiplets.

The renormalizable superpotential is
\begin{equation}
W=W_{\rm Higgs}+W_{\rm Yukawa},
\end{equation}
with
\begin{align}
W_{\rm Higgs}
&=
\frac{1}{3}\lambda_\Sigma {\rm Tr}\Sigma^3
+\frac{1}{2}m_\Sigma {\rm Tr}\Sigma^2
+\lambda_H \bar H\Sigma H
+m_H\bar H H,
\label{superpotentialHiggs}
\\
W_{\rm Yukawa}
&=
\frac{1}{4}h^{ij}
\epsilon_{\hat a\hat b\hat c\hat d\hat e}
\Psi_i^{\hat a\hat b}
\Psi_j^{\hat c\hat d}
H^{\hat e}
-\sqrt{2}f^{ij}
\Psi_i^{\hat a\hat b}
\Phi_{j\hat a}
\bar H_{\hat b},
\label{superpotentialYukawa}
\end{align}
where \(\hat a,\hat b,\ldots=1,\ldots,5\) are \(\mathrm{SU}(5)\) indices and
\(\epsilon_{12345}=1\). We assume \(R\)-parity conservation, which forbids superpotential terms such as \(\Psi\Phi\Phi\) and \(H\Phi\).

The Yukawa matrices \(h^{ij}\) and \(f^{ij}\) in
Eq.~\eqref{superpotentialYukawa} contain redundant flavor degrees of
freedom, part of which can be removed by redefining the matter
superfields \(\Psi_i\) and \(\Phi_i\). Following Ref.~\cite{Hisano:1992jj},
we choose a basis in which
\begin{align}
h^{ij}
&=
e^{i\varphi_i}\,
\delta_{ij}\,
f_{u_i}(\mu_{\rm GUT}),
\label{eq:hij}
\\
f^{ij}
&=
V^*_{ij}\,
f_{d_j}(\mu_{\rm GUT}),
\end{align}
where \(f_{u_i}(\mu_{\rm GUT})\) and
\(f_{d_i}(\mu_{\rm GUT})\) denote the up-type and down-type Yukawa
couplings, respectively, evaluated at the GUT scale
\(\mu_{\rm GUT}\), and \(V_{ij}\) is the
Cabibbo--Kobayashi--Maskawa (CKM) matrix. The phases \(\varphi_i\)
satisfy
\begin{equation}
\sum_i \varphi_i = 0,
\end{equation}
so that two independent phases remain~\cite{Ellis:1979hy}. In this basis, the MSSM matter superfields are embedded into the
\(\mathrm{SU}(5)\) multiplets as
\begin{align}
\Psi_i
&\ni
\left\{
Q_i,\,
e^{-i\varphi_i}\bar U_i,\,
V_{ij}\bar E_j
\right\},
\\
\Phi_i
&\ni
\left\{
\bar D_i,\,
L_i
\right\}.
\end{align}
The phases \(\varphi_i\), which do not affect the SM Yukawa interactions, enter the couplings of the color-triplet Higgs fields and hence can affect the dimension-five proton-decay amplitudes.

The adjoint Higgs develops a vacuum expectation value
\begin{equation}
\langle\Sigma\rangle
=
V\,{\rm diag}(2,2,2,-3,-3),
\end{equation}
which breaks \(\mathrm{SU}(5)\) down to the SM gauge group. Requiring supersymmetry to remain unbroken at this stage gives
\begin{equation}
V=\frac{m_\Sigma}{\lambda_\Sigma}.
\end{equation}
The masses of the physical components of the adjoint Higgs and the color-triplet Higgs are then
\begin{equation}
M_\Sigma
\equiv
M_{\Sigma_8}
=
M_{\Sigma_3}
=
\frac{5}{2}\lambda_\Sigma V,
\qquad
M_{\Sigma_{24}}
=
\frac{1}{2}\lambda_\Sigma V,
\qquad
M_{H_C}
=
5\lambda_H V.
\label{eq:GUTmasses}
\end{equation}
The MSSM \(\mu\) parameter is
\begin{equation}
\mu=m_H-3\lambda_H V.
\end{equation}
Thus, the familiar doublet--triplet splitting requires a fine-tuning between \(m_H\) and \(3\lambda_H V\).

The gauge interactions of the adjoint Higgs give the \(X\)-boson mass
\begin{equation}
M_X=5\sqrt{2}\,g_{\rm GUT}V,
\label{eq:MX}
\end{equation}
where \(g_{\rm GUT}\) is the unified gauge coupling. The \((\mathbf{3},\mathbf{2})\) and \((\overline{\mathbf{3}},\mathbf{2})\) components of \(\Sigma\) are absorbed into the massive \(X\) gauge multiplets and therefore do not appear as physical states.

\subsection{Gauge coupling unification and threshold corrections}
\label{sec:gcu}

The masses of the superheavy particles are crucial for proton decay. In particular, \(X\)-boson exchange induces dimension-six proton decay, whereas color-triplet Higgs exchange generates the dimension-five baryon-number-violating operators characteristic of supersymmetric GUTs~\cite{Sakai:1981pk, Weinberg:1981wj}. We estimate these masses using the GUT-scale threshold corrections to the gauge couplings, following Refs.~\cite{Hisano:1992mh,Hisano:1992jj,Hisano:2013cqa}.

We work in the \(\overline{\rm DR}\) scheme~\cite{Siegel:1979wq}. At a matching scale \(\mu_{\rm GUT}\) around the masses of the superheavy particles, the one-loop matching conditions for the gauge couplings are~\cite{Weinberg:1980wa,Hall:1980kf}
\begin{align}
\frac{1}{g_1^2(\mu_{\rm GUT})}
&=
\frac{1}{g_{\rm GUT}^2(\mu_{\rm GUT})}
+\frac{1}{8\pi^2}
\left[
\frac{2}{5}
\ln\frac{\mu_{\rm GUT}}{M_{H_C}}
-10\ln\frac{\mu_{\rm GUT}}{M_X}
\right],
\nonumber\\
\frac{1}{g_2^2(\mu_{\rm GUT})}
&=
\frac{1}{g_{\rm GUT}^2(\mu_{\rm GUT})}
+\frac{1}{8\pi^2}
\left[
2\ln\frac{\mu_{\rm GUT}}{M_\Sigma}
-6\ln\frac{\mu_{\rm GUT}}{M_X}
\right],
\nonumber\\
\frac{1}{g_3^2(\mu_{\rm GUT})}
&=
\frac{1}{g_{\rm GUT}^2(\mu_{\rm GUT})}
+\frac{1}{8\pi^2}
\left[
\ln\frac{\mu_{\rm GUT}}{M_{H_C}}
+3\ln\frac{\mu_{\rm GUT}}{M_\Sigma}
-4\ln\frac{\mu_{\rm GUT}}{M_X}
\right]\,,
\label{eq:GUTmatching}
\end{align}
where we use the conventional GUT normalization for \(g_1\): \(g_1 \equiv \sqrt{5/3}\, g'\), with \(g'\) the U(1)$_Y$ gauge coupling. Since we use the \(\overline{\rm DR}\) scheme, there are no scale-independent constant terms in these matching conditions. It should be noted that the matching conditions in Eq.~\eqref{eq:GUTmatching} can be significantly modified in the presence of Planck-suppressed higher-dimensional operators. In particular, the dimension-five operator
\begin{equation}
W_{\rm eff}^{\Delta g}
=
\frac{c}{M_P}\,
{\rm Tr}\!\left[\Sigma {\cal W}{\cal W}\right],
\end{equation}
where \({\cal W}\) denotes the gauge field-strength chiral superfield of \(\mathrm{SU}(5)\), induces non-universal corrections to the gauge couplings after \(\Sigma\) acquires its vacuum expectation value. These corrections can be comparable to the GUT-scale threshold corrections and can therefore substantially alter the inferred values of the superheavy masses~\cite{Tobe:2003yj}. In the present analysis, we do not include this operator and restrict ourselves to the renormalizable minimal supersymmetric \(\mathrm{SU}(5)\) setup. For discussions of the impact of such Planck-suppressed operators on gauge coupling matching and proton decay, see, e.g., Refs.~\cite{Ellis:2016tjc, Ellis:2019fwf, Evans:2021hyx, Ellis:2026tgb}.

Taking appropriate linear combinations eliminates the unified gauge coupling \(g_{\rm GUT}\), yielding
\begin{align}
\frac{3}{g_2^2(\mu_{\rm GUT})}
-\frac{2}{g_3^2(\mu_{\rm GUT})}
-\frac{1}{g_1^2(\mu_{\rm GUT})}
&=
-\frac{3}{10\pi^2}
\ln\left(
\frac{\mu_{\rm GUT}}{M_{H_C}}
\right),
\label{eq:MHccondition}
\\
\frac{5}{g_1^2(\mu_{\rm GUT})}
-\frac{3}{g_2^2(\mu_{\rm GUT})}
-\frac{2}{g_3^2(\mu_{\rm GUT})}
&=
-\frac{9}{2\pi^2}
\ln\left(
\frac{\mu_{\rm GUT}}{M_{\rm GUT}}
\right),
\label{eq:MGUTcondition}
\end{align}
where we have defined
\begin{equation}
M_{\rm GUT}
\equiv
\left(M_X^2M_\Sigma\right)^{1/3}.
\label{eq:MGUTdef}
\end{equation}
These two relations play a central role in our analysis. Once the low-energy gauge couplings are evolved to \(\mu_{\rm GUT}\), including the threshold corrections associated with the supersymmetric spectrum, they determine \(M_{H_C}\) and the combination \(M_{\rm GUT}=(M_X^2M_\Sigma)^{1/3}\).

To illustrate how the GUT-scale masses depend on the supersymmetric mass spectrum, it is useful to derive approximate analytic relations at the one-loop level. Solving the one-loop renormalization group equations (RGEs) for the gauge couplings and including the threshold corrections from the supersymmetric particles, Eqs.~\eqref{eq:MHccondition} and \eqref{eq:MGUTcondition} can be expressed as
\begin{align}
\frac{3}{\alpha_2(m_Z)}
-\frac{2}{\alpha_3(m_Z)}
-\frac{1}{\alpha_1(m_Z)}
&=
\frac{1}{2\pi}
\biggl[
\frac{12}{5}\ln\biggl(\frac{M_{H_C}}{m_Z}\biggr)\nonumber\\
&-\frac{8}{5}\ln\biggl(\frac{M_{\widetilde H}}{m_Z}\biggr)
-\frac{2}{5}\ln\biggl(\frac{M_H}{m_Z}\biggr)
+4\ln\biggl(\frac{M_3}{M_2}\biggr)
\biggr],
\label{mhc}
\\
\frac{5}{\alpha_1(m_Z)}
-\frac{3}{\alpha_2(m_Z)}
-\frac{2}{\alpha_3(m_Z)}
&=
\frac{1}{2\pi}
\biggl[
36\ln\biggl(\frac{M_{\rm GUT}}{m_Z}\biggr)
+4\ln\biggl(\frac{M_2}{m_Z}\biggr)
+4\ln\biggl(\frac{M_3}{m_Z}\biggr)
\biggr],
\label{mgut}
\end{align}
where \(\alpha_i\equiv g_i^2/(4\pi)\). Here, \(m_Z\) is the \(Z\)-boson mass, \(M_3\) and \(M_2\) denote the gluino and wino masses, respectively, \(M_{\widetilde H}\) is the Higgsino mass, and \(M_H\) is the mass of the heavy Higgs doublet. For simplicity, we have taken the sfermion masses to be degenerate. At this order, their common mass does not appear in the above combinations. Consequently, the determination of these GUT-scale mass parameters is largely insensitive to the choice of the sfermion mass scale.

These relations make the qualitative dependence on the supersymmetric spectrum transparent. Equation~\eqref{mgut} shows that raising the overall gaugino mass scale lowers \(M_{\rm GUT}\). On the other hand, Eq.~\eqref{mhc} shows that \(M_{H_C}\) increases as the Higgsino mass is raised, whereas it decreases for a larger hierarchy \(M_3/M_2\). The latter dependence will be particularly important in the heavy-Higgsino scenario considered below, since a gluino substantially heavier than the wino can bring \(M_{H_C}\) back toward the conventional supersymmetric GUT scale even when the Higgsino is very heavy.

Equations~\eqref{mhc} and \eqref{mgut} are intended only to provide a simple analytic understanding of these dependences. In the numerical analysis below, we determine the gauge-coupling evolution using the two-loop RGEs, together with the appropriate threshold corrections.

It is worth emphasizing that the second relation does not determine \(M_X\) itself. Using Eqs.~\eqref{eq:GUTmasses} and \eqref{eq:MX}, one finds
\begin{equation}
\frac{M_\Sigma}{M_X}
=
\frac{\lambda_\Sigma}{2\sqrt{2}\,g_{\rm GUT}}.
\end{equation}
Consequently, for a sufficiently small adjoint-Higgs self-coupling \(\lambda_\Sigma\), \(M_\Sigma\) can be substantially smaller than \(M_X\) while keeping \(M_{\rm GUT}\) fixed. Equivalently,
\begin{equation}
M_X
=
\left(
\frac{2\sqrt{2}\,g_{\rm GUT}}{\lambda_\Sigma}
\right)^{1/3}
M_{\rm GUT}.
\label{eq:MXMGUT}
\end{equation}
The \(X\)-boson mass can therefore be significantly higher than the scale \(M_{\rm GUT}\) inferred directly from gauge coupling unification. This observation will be important in Sec.~\ref{sec:protondec}, where we show that dimension-six proton decay can remain sufficiently suppressed even for the heavy supersymmetric spectrum.

By contrast, \(M_{H_C}\) itself is directly determined by the first combination in Eq.~\eqref{eq:MHccondition}. Its value is particularly sensitive to the supersymmetric threshold corrections, and hence to the hierarchy among the Higgsino, wino, and gluino masses. As we will see below, the heavy-Higgsino spectrum of interest can yield \(M_{H_C}\) close to the conventional supersymmetric GUT scale, \(M_{H_C}\sim 10^{16}~{\rm GeV}\), provided that the gluino is sufficiently heavier than the wino. This feature has direct implications for the dimension-five proton decay rate.

\section{Proton decay for the heavy Higgsino scenario}
\label{sec:protondec}

\subsection{Heavy Higgsino scenario for the LZ event}

We begin by summarizing the supersymmetric mass spectrum motivated by the heavy-Higgsino interpretation of the LZ event. Ref.~\cite{Langhoff:2026ujr} performed a likelihood analysis of the Higgsino mass \(M_{\widetilde H}\) and the mass splitting \(\delta m\) between the two neutral Higgsino states, taking into account the \(248~{\rm keV}\) recoil event, the absence of events in the high-energy sideband, and the constraint from Higgsino capture in the Sun. While the conventional thermal Higgsino with \(M_{\widetilde H}\simeq 1.1~{\rm TeV}\) is in tension with these observations, a heavy-Higgsino region remains viable. In particular, the parameter region most consistent with the combined constraints is \(M_{\widetilde H}\sim 10^5\text{--}10^6~{\rm GeV}\) and \(\delta m\sim 330\text{--}480~{\rm keV}\), with some dependence on the assumed dark-matter velocity distribution. Representative best-fit points on the heavy branch have \(M_{\widetilde H}\simeq (1.0\text{--}3.2)\times10^5~{\rm GeV}\) and \(\delta m\simeq 365\text{--}449~{\rm keV}\). In what follows, we take this region as the target spectrum and examine its implications for supersymmetric grand unification and proton decay.

\begin{figure}
  \centering
  \includegraphics[width=0.55\linewidth]{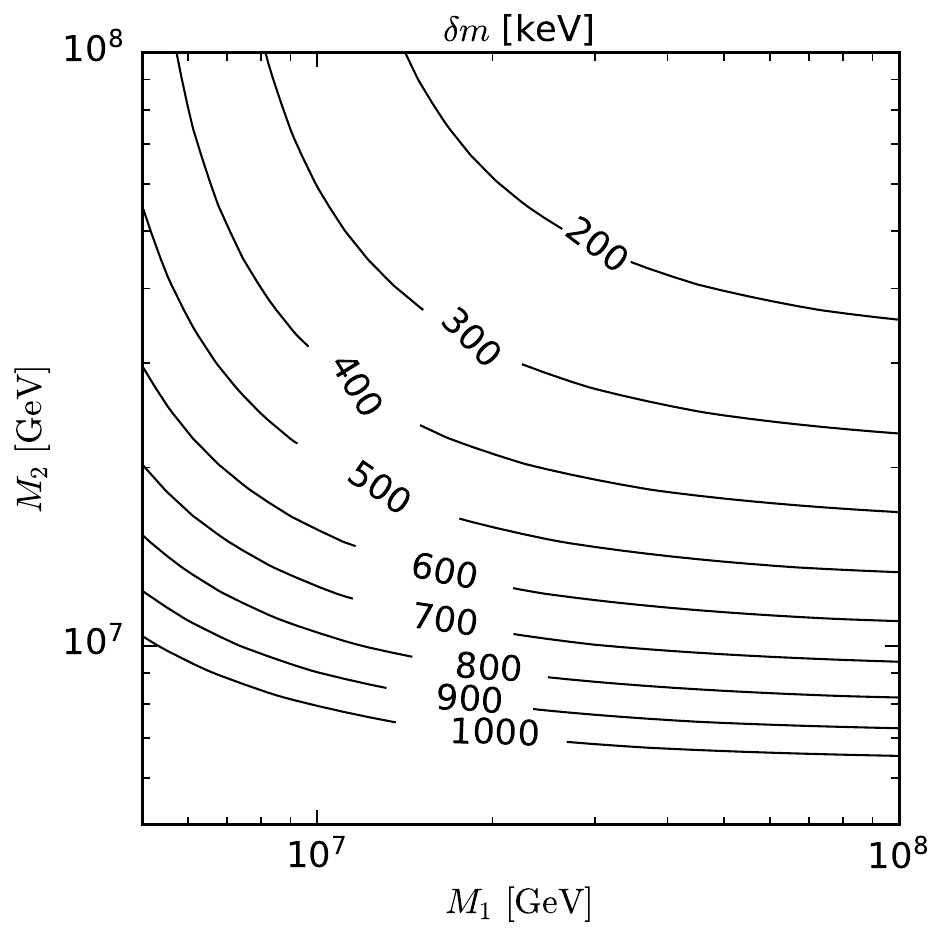}
  \caption{
  Contours of the mass splitting $\delta m$ between the
  two neutral Higgsino states in the $M_1$--$M_2$ plane. The contour labels are given in units of keV. 
  }
  \label{fig:M1M2delta}
\end{figure}

The neutral-Higgsino mass splitting also determines the characteristic scale of the gaugino masses.  In the limit $|M_1|,|M_2|\gg |\mu|,m_Z$, integrating out the bino and wino gives,
at leading order,
\begin{equation}
\delta m
\equiv
m_{\widetilde\chi^0_2}-m_{\widetilde\chi^0_1}
\simeq
m_Z^2
\left|
\frac{s_W^2}{M_1}
+
\frac{c_W^2}{M_2}
\right|,
\label{eq:higgsino-splitting}
\end{equation}
where $s_W\equiv\sin\theta_W$ and $c_W\equiv\cos\theta_W$, with \(\theta_W\) the weak mixing angle, and the Higgsino mass is given by \(M_{\widetilde H}\simeq |\mu|\). For illustration, Fig.~\ref{fig:M1M2delta} shows contours of $\delta m$ in the $M_1$--$M_2$ plane.  The mass splitting favored by the heavy-Higgsino interpretation of the LZ event,
$\delta m\simeq 330$--$480~{\rm keV}$, points to a
characteristic gaugino mass scale of order $10^7~{\rm GeV}$.  In
particular, for $M_1\simeq M_2$, Eq.~\eqref{eq:higgsino-splitting}
reduces to
\begin{equation}
\delta m
\simeq
\frac{m_Z^2}{M_2}
\simeq
420~{\rm keV}
\left(\frac{2.0\times10^7~{\rm GeV}}{M_2}\right)\,.
\end{equation}
Since the wino contribution in Eq.~\eqref{eq:higgsino-splitting} is weighted by $c_W^2$, the splitting is more sensitive to $M_2$ than to $M_1$.  This behavior is also evident in Fig.~\ref{fig:M1M2delta}, where the contours approach approximately horizontal lines for relatively large $M_1$. Thus, the few-hundred-keV neutral-Higgsino splitting required by the LZ interpretation robustly suggests electroweak gaugino masses around $10^7~{\rm GeV}$, while still allowing a sizable hierarchy between $M_1$ and $M_2$.

\subsection{Proton decay in minimal supersymmetric SU(5)}

In minimal supersymmetric \(\mathrm{SU}(5)\), proton decay is induced mainly through two classes of baryon-number-violating interactions. The exchange of the superheavy \(\mathrm{SU}(5)\) gauge bosons generates dimension-six operators, leading predominantly to decay modes such as \(p\to \pi^0 e^+\). On the other hand, the exchange of the color-triplet Higgs superfields generates the Sakai--Yanagida--Weinberg dimension-five operators~\cite{Sakai:1981pk,Weinberg:1981wj}, which, after being dressed by supersymmetric particles, typically give rise to the \(p\to K^+\bar{\nu}\) mode~\cite{Dimopoulos:1981dw, Ellis:1981tv}. 

We first consider proton decay induced by the exchange of the
\(\mathrm{SU}(5)\) gauge bosons. Integrating out the \(X\) bosons at the
GUT scale generates baryon-number-violating dimension-six operators.
In superspace notation, the corresponding effective Lagrangian can be
written as
\begin{equation}
{\cal L}_{6}^{\rm eff}
=
C_{6(1)}^{ijkl}{\cal O}^{6(1)}_{ijkl}
+
C_{6(2)}^{ijkl}{\cal O}^{6(2)}_{ijkl} + \mathrm{h.c.},
\label{eq:dim6eff}
\end{equation}
where
\begin{align}
{\cal O}^{6(1)}_{ijkl}
&=
\int d^2\theta d^2\bar{\theta}\,
\epsilon_{abc}\epsilon_{\alpha\beta}
\bigl(\bar U_i^\dagger\bigr)^a
\bigl(\bar D_j^\dagger\bigr)^b
e^{-\frac{2}{3}g' B}
\bigl(e^{2g_3G}Q_k^\alpha\bigr)^c
L_l^\beta,
\label{eq:dim6op1}
\\
{\cal O}^{6(2)}_{ijkl}
&=
\int d^2\theta d^2\bar{\theta}\,
\epsilon_{abc}\epsilon_{\alpha\beta}
Q_i^{a\alpha}Q_j^{b\beta}
e^{\frac{2}{3}g' B}
\bigl(e^{-2g_3G}\bar U_k^\dagger\bigr)^c
\bar E_l^\dagger.
\label{eq:dim6op2}
\end{align}
Here, \(a,b,c\) and \(\alpha,\beta\) denote the
\(\mathrm{SU}(3)_C\) and \(\mathrm{SU}(2)_L\) indices, respectively.

At the matching scale, the tree-level exchange of the \(X\) bosons
gives
\begin{align}
C^{ijkl}_{6(1)}
&=
-\frac{g_{\rm GUT}^2}{M_X^2}\,
e^{i\varphi_i}\,
\delta^{ik}\delta^{jl},
\nonumber\\
C^{ijkl}_{6(2)}
&=
-\frac{g_{\rm GUT}^2}{M_X^2}\,
e^{i\varphi_i}\,
\delta^{ik}(V^*)^{jl},
\label{eq:dim6gutmatch}
\end{align}
where \(V\) is the CKM matrix and the phases \(\varphi_i\) are those introduced in the GUT-scale Yukawa couplings above. The proton-decay amplitudes generated by these operators are independent of \(\varphi_i\), since these phases enter only as overall phase factors.

On the other hand, integrating out the color-triplet Higgs superfields generates baryon-number-violating dimension-five operators~\cite{Sakai:1981pk,Weinberg:1981wj}. The corresponding effective Lagrangian is given by
\begin{equation}
{\cal L}_5^{\rm eff}
=
C^{ijkl}_{5L}{\cal O}^{5L}_{ijkl}
+
C^{ijkl}_{5R}{\cal O}^{5R}_{ijkl}
+\mathrm{h.c.},
\label{eq:dim5eff}
\end{equation}
where
\begin{align}
{\cal O}^{5L}_{ijkl}
&\equiv
\int d^2\theta\,
\frac{1}{2}\epsilon_{abc}
(Q_i^a\cdot Q_j^b)
(Q_k^c\cdot L_l),
\nonumber\\
{\cal O}^{5R}_{ijkl}
&\equiv
\int d^2\theta\,
\epsilon^{abc}
\bar U_{ia}\bar E_j
\bar U_{kb}\bar D_{lc}.
\end{align}
At the GUT scale, the corresponding Wilson coefficients are
\begin{align}
C^{ijkl}_{5L}(\mu_{\rm GUT})
&=
\frac{1}{M_{H_C}}\,
f_{u_i}\,
e^{i\varphi_i}\,
\delta^{ij}
V^*_{kl}\,
f_{d_l},
\nonumber\\
C^{ijkl}_{5R}(\mu_{\rm GUT})
&=
\frac{1}{M_{H_C}}\,
f_{u_i}\,
V_{ij}V^*_{kl}\,
f_{d_l}\,
e^{-i\varphi_k}.
\label{eq:wilson5}
\end{align}
Unlike the dimension-six contribution, the dimension-five operators depend explicitly on the GUT phases \(\varphi_i\), and hence the proton decay amplitudes can in general be sensitive to these phases. Owing to the antisymmetric contraction of the color indices, these operators necessarily involve quarks from at least two generations. As a result, the dominant decay modes typically contain a strange quark, with \(p\to K^+\bar{\nu}\) providing the most important channel in minimal supersymmetric \(\mathrm{SU}(5)\).

Below the supersymmetry-breaking scale, the baryon-number-violating interactions are described by four-fermion operators~\cite{Weinberg:1979sa,Wilczek:1979hc,Abbott:1980zj, Hamaguchi:2024ewe}. For the dimension-six operators induced by \(X\)-boson exchange, the matching onto the corresponding four-fermion operators is straightforward. The situation is different for the dimension-five operators generated by color-triplet Higgs exchange. In this case, the effective operators in Eq.~\eqref{eq:dim5eff} contain two fermions and two sfermions and must be converted into four-fermion operators through one-loop diagrams involving wino or Higgsino exchange~\cite{Goto:1998qg}, unless sizable flavor violation is present in the sfermion mass matrices~\cite{Nagata:2013sba}. Through the loop functions arising in this matching, the resulting Wilson coefficients depend on the wino mass, the Higgsino mass, and the sfermion masses.

We now summarize the procedure used in our numerical calculation of the proton decay rates. For the dimension-six operators, we evolve the Wilson coefficients from the GUT scale down to the supersymmetry-breaking scale using the one-loop RGEs given in Ref.~\cite{Munoz:1986kq}.\footnote{The two-loop RGEs of the dimension-six proton-decay operators have been calculated in Ref.~\cite{Hisano:2013ege}, and the two-loop corrections were found to be numerically small.} For the dimension-five operators, on the other hand, the non-renormalization theorem implies that the evolution of their Wilson coefficients is determined by the wave-function renormalization of the chiral superfields entering the operators; we include these effects at the one-loop level. At the supersymmetry-breaking scale, the dimension-five operators are matched onto four-fermion operators through the one-loop dressing diagrams, for which we use the matching conditions given in Ref.~\cite{Hisano:2013exa}. We then evolve the Wilson coefficients of the four-fermion operators from the supersymmetry-breaking scale down to the electroweak scale using the one-loop RGEs of Ref.~\cite{Alonso:2014zka}. Below the electroweak scale, we include the two-loop QCD renormalization effects following Ref.~\cite{Nihei:1994tx}. Finally, for the hadronic matrix elements, we use the results obtained from lattice QCD simulations in Ref.~\cite{Yoo:2021gql}. Further details of the calculation can be found in Ref.~\cite{Hamaguchi:2022yjo}.

\subsection{Results}
\label{sec:results}

\begin{figure}
  \centering
  \begin{subfigure}[t]{0.48\textwidth}
    \centering
    \includegraphics[width=\linewidth]{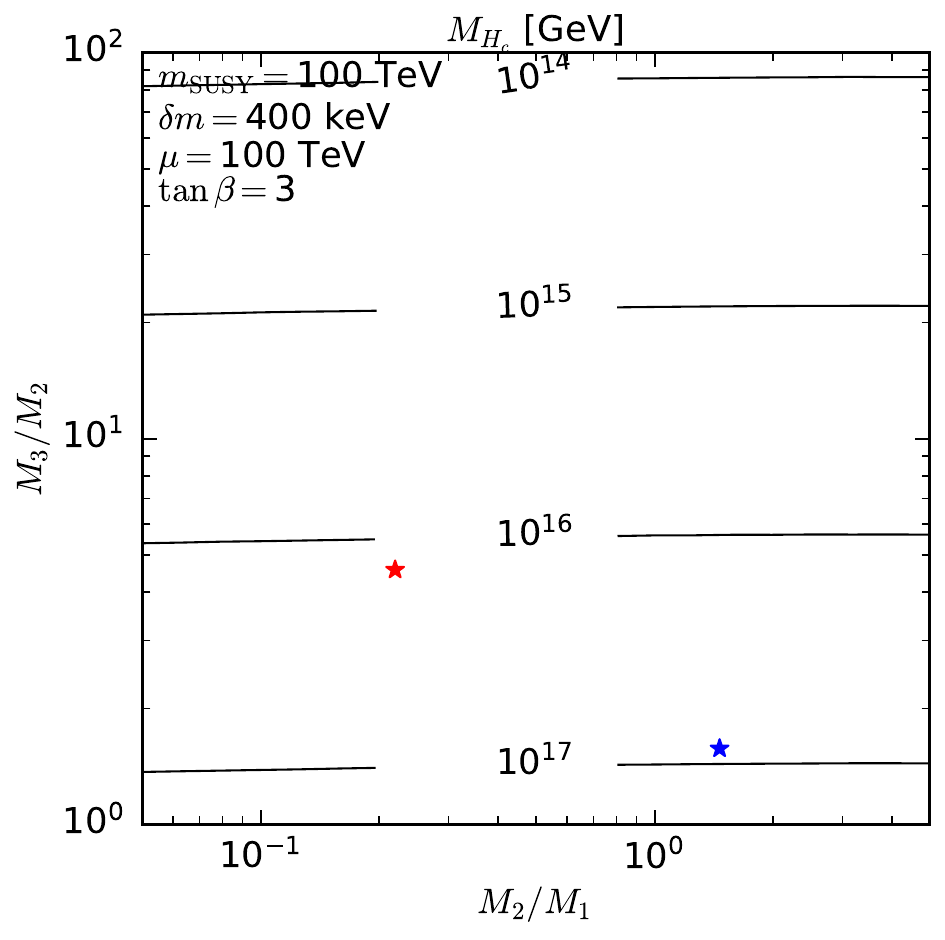}
    \caption{
    Color-triplet Higgs mass \(M_{H_C}\).
    }
    \label{fig:gutmasses_mhc}
  \end{subfigure}
  \hfill
  \begin{subfigure}[t]{0.48\textwidth}
    \centering
    \includegraphics[width=\linewidth]{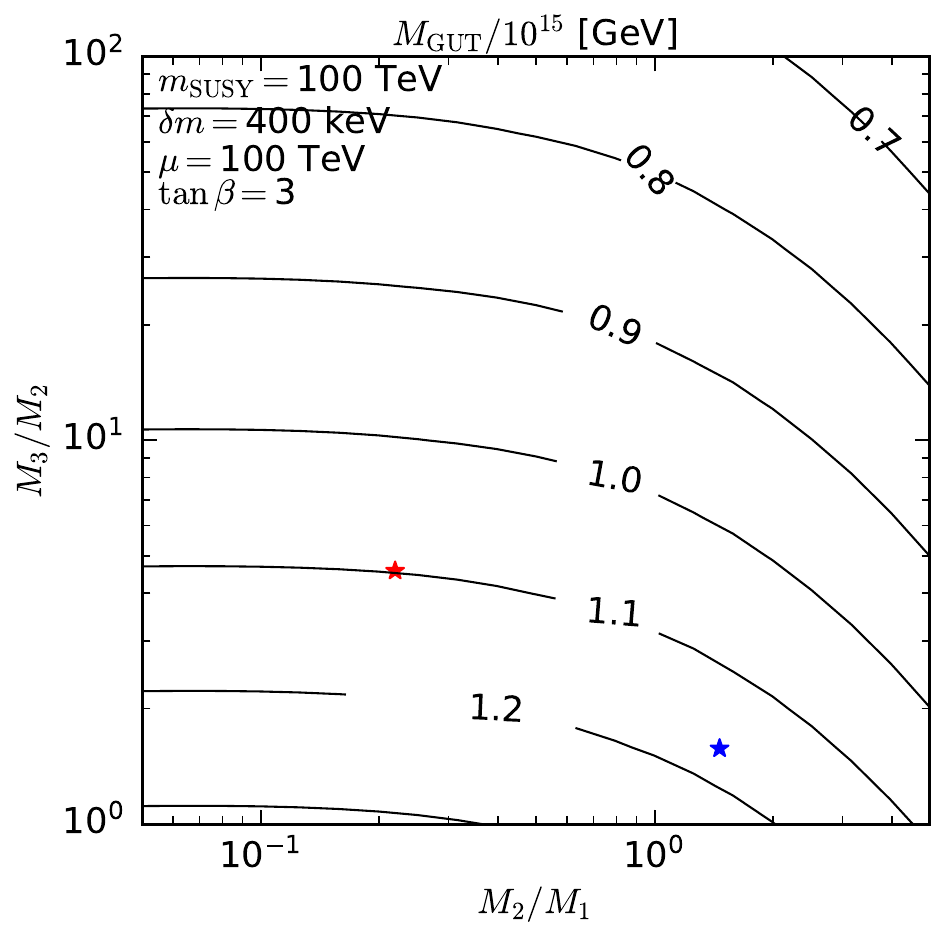}
    \caption{
    \(M_{\rm GUT}\equiv(M_X^2M_\Sigma)^{1/3}\).
    }
    \label{fig:gutmasses_mgut}
  \end{subfigure}
  \caption{
  GUT-scale mass parameters determined from gauge coupling unification in the \(M_2/M_1\)--\(M_3/M_2\) plane. We take the sfermion masses and the heavy Higgs-boson mass to be \(100~{\rm TeV}\), \(\mu=100~{\rm TeV}\), and \(\tan\beta=3\), and determine  the overall gaugino mass scale by imposing
  \(\delta m=400~{\rm keV}\). The red and blue stars indicate the gaugino-mass ratios expected from anomaly mediation~\cite{Randall:1998uk,Giudice:1998xp} and gaugino-mass unification, respectively.
  }
  \label{fig:gutmasses}
\end{figure}

We first determine the GUT-scale mass parameters \(M_{H_C}\) and \(M_{\rm GUT}\) using the gauge-coupling unification conditions discussed in Sec.~\ref{sec:gcu}. Figure~\ref{fig:gutmasses} shows the resulting values of \(M_{H_C}\) and \(M_{\rm GUT}\) in the \(M_2/M_1\)--\(M_3/M_2\) plane. We take the sfermion masses \(m_{\widetilde{f}}\) and the heavy Higgs-boson mass \(M_H\) to be \(m_{\rm SUSY}\equiv m_{\widetilde{f}}=M_H=100~{\rm TeV}\), while the Higgsino mass is fixed to \(\mu=100~{\rm TeV}\), and set \(\tan\beta=3\).\footnote{We take \(\tan\beta=3\) since this value approximately reproduces the observed SM Higgs-boson mass for the sfermion mass scale considered here~\cite{Bagnaschi:2014rsa}.} As discussed in Sec.~\ref{sec:gcu}, the determination of \(M_{H_C}\) and \(M_{\rm GUT}\) is largely insensitive to the choice of \(m_{\widetilde f}\). For each value of \(M_2/M_1\), the overall gaugino mass scale is determined by requiring the neutral-Higgsino mass splitting to be \(\delta m =400~{\rm keV}\). We use the two-loop RGEs for the gauge couplings and include the one-loop supersymmetric threshold corrections. The electroweak-scale threshold corrections are incorporated following Ref.~\cite{Buttazzo:2013uya}. The red stars indicate the gaugino-mass ratios predicted by anomaly mediation~\cite{Randall:1998uk,Giudice:1998xp}, for which the gaugino masses are given by
\begin{equation}
M_i
=
\frac{b_i g_i^2}{16\pi^2}m_{3/2},
\label{eq:amsb-gaugino}
\end{equation}
where \(m_{3/2}\) is the gravitino mass and \(b_i\) are the one-loop beta-function coefficients. In the MSSM,
\begin{equation}
(b_1,b_2,b_3)
=
\left(\frac{33}{5},1,-3\right).
\end{equation}
The blue stars, on the other hand, indicate the gaugino-mass ratios expected from gaugino-mass unification, as is often realized in gravity-mediated supersymmetric GUTs. In this case, the gaugino masses are universal at the GUT scale, and the ratios \(M_i/g_i^2\) remain universal and renormalization-scale independent at the one-loop level.

The behavior of \(M_{H_C}\) in the left panel can be understood from the approximate one-loop relation in Eq.~\eqref{mhc}. The color-triplet Higgs mass depends strongly on the ratio \(M_3/M_2\), while its dependence on \(M_2/M_1\), which arises from higher-order running and threshold effects, is comparatively weak. In particular, increasing \(M_3/M_2\) lowers the value of \(M_{H_C}\). Thus, although the large Higgsino mass tends to raise \(M_{H_C}\), this effect can be compensated by taking the gluino sufficiently heavier than the wino. For \(M_3/M_2 \gg 1\), \(M_{H_C}\) can remain close to the conventional supersymmetric GUT scale, \(M_{H_C}\sim10^{16}~{\rm GeV}\).

The right panel shows that \(M_{\rm GUT}\) is typically of 
\(\mathcal{O}(10^{15})~{\rm GeV}\) for the heavy gaugino spectrum considered here. This reduction compared with the conventional supersymmetric unification scale is expected from Eq.~\eqref{mgut}: increasing the gaugino mass scale lowers \(M_{\rm GUT}\). In contrast to \(M_{H_C}\), \(M_{\rm GUT}\) is sensitive not only to \(M_3/M_2\) but also to \(M_2/M_1\), since the latter affects the absolute wino mass once the Higgsino mass splitting is fixed.

\begin{figure}
  \centering
  \includegraphics[width=0.55\linewidth]
  {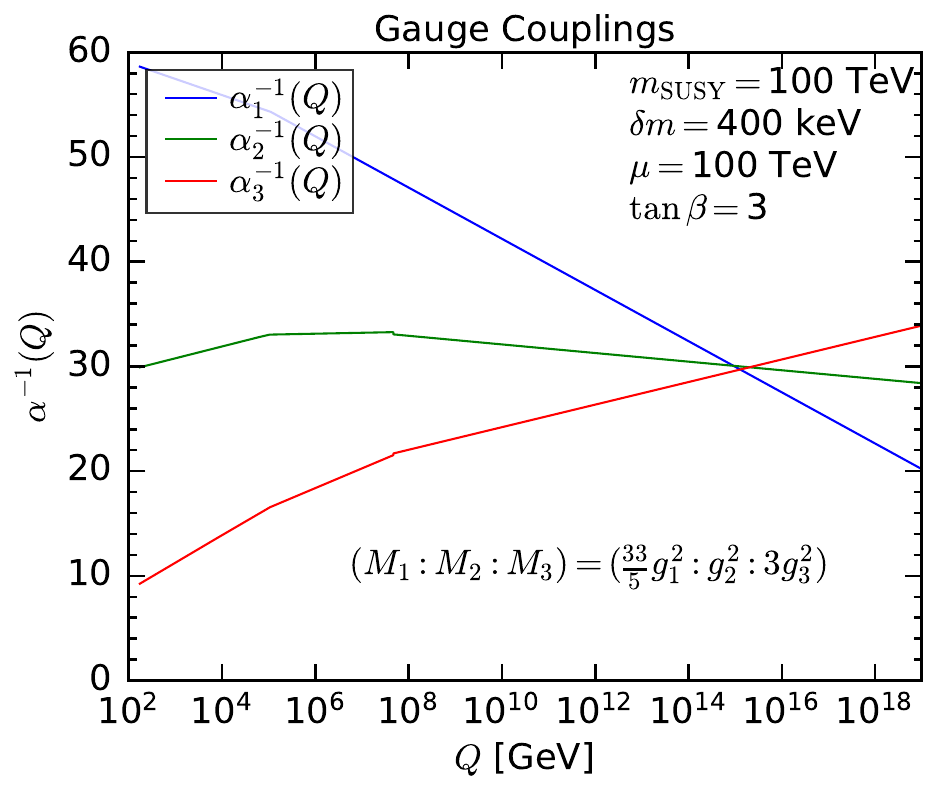}
  \caption{
  Renormalization-group evolution of the inverse gauge couplings
  \(\alpha_i^{-1}(Q)\) for the gaugino-mass ratios predicted by
  anomaly mediation. We take \(m_{\rm SUSY}=100~{\rm TeV}\), \(\mu=100~{\rm TeV}\), \(\tan\beta=3\), and determine the overall gaugino mass scale by imposing \(\delta m=400~{\rm keV}\). 
  }
  \label{fig:gaugecoupling_amsb}
\end{figure}

It is also instructive to discuss the quality of gauge coupling
unification in the present scenario. Within the framework described above, a necessary and sufficient condition for the three gauge couplings evolved from low energies to meet at a single scale is that there exists a matching scale \(\mu_{\rm GUT}\) at which the right-hand sides of both Eqs.~\eqref{eq:MHccondition} and \eqref{eq:MGUTcondition} vanish simultaneously. This requires \(M_{H_C}=M_{\rm GUT}\). Therefore, the closer \(M_{H_C}\) and \(M_{\rm GUT}\) are to each other, the more closely the three gauge couplings approach one another at high energies. In the present case, increasing \(M_3/M_2\) lowers \(M_{H_C}\) while leaving \(M_{\rm GUT}\) comparatively less affected, bringing the two GUT-scale mass parameters closer to each other. With a sizable hierarchy between the gluino and wino masses, therefore, high-quality gauge coupling unification can still be achieved even when the overall gaugino mass scale is very high. This is illustrated explicitly in Fig.~\ref{fig:gaugecoupling_amsb}, where we show the evolution of the three SM gauge couplings for the gaugino-mass ratios predicted by anomaly mediation. We take \(\delta m=400~{\rm keV}\), \(\mu=100~{\rm TeV}\), \(m_{\rm SUSY}=100~{\rm TeV}\), and \(\tan\beta=3\).  The overall gaugino mass scale is fixed by the Higgsino mass-splitting condition. As can be seen from the figure, the three gauge couplings approach one another very closely around the GUT scale, demonstrating explicitly that the heavy-gaugino spectrum considered here can preserve a high degree of gauge coupling unification.

\begin{figure}
  \centering
  \includegraphics[width=0.55\linewidth]{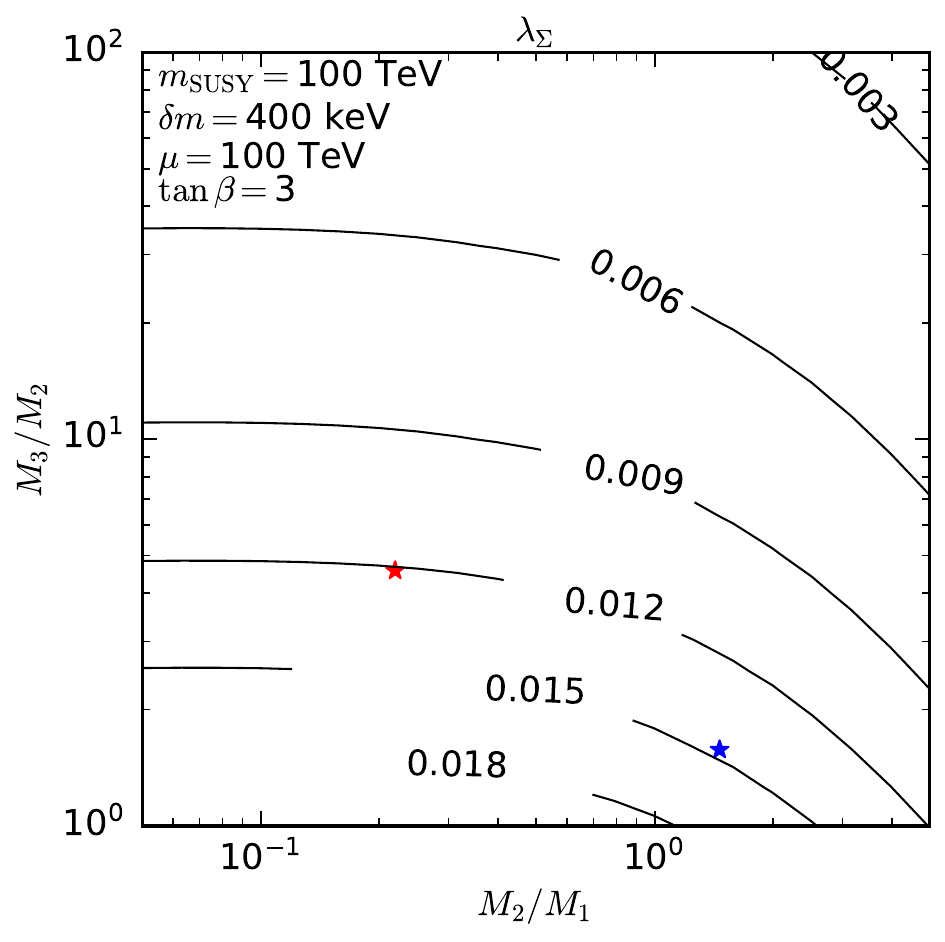}
  \caption{
  Upper bound on the adjoint-Higgs self-coupling \(\lambda_\Sigma\) required for the \(p\to\pi^0 e^+\) lifetime induced by \(\mathrm{SU}(5)\) gauge-boson exchange to satisfy the current experimental limit, \(\tau (p \to\pi^0 e^+) > 2.4\times 10^{34}~\mathrm{years}\)~\cite{Super-Kamiokande:2020wjk}. We take the sfermion masses and the heavy Higgs-boson mass to be \(100~{\rm TeV}\), \(\mu=100~{\rm TeV}\), and \(\tan\beta=3\), and determine the overall gaugino mass scale by imposing  \(\delta m=400~{\rm keV}\). The red and blue stars indicate the gaugino-mass ratios expected from anomaly mediation~\cite{Randall:1998uk,Giudice:1998xp} and gaugino-mass unification, respectively.
  }
  \label{fig:lambdasigma_bound}
\end{figure}

We next examine the constraint from dimension-six proton decay mediated by the \(\mathrm{SU}(5)\) gauge bosons. As discussed in Sec.~\ref{sec:gcu}, for a fixed value of \(M_{\rm GUT}\), the \(X\)-boson mass can be raised by decreasing the adjoint-Higgs self-coupling \(\lambda_\Sigma\). Figure~\ref{fig:lambdasigma_bound} shows the upper bound on \(\lambda_\Sigma\) required for the \(p\to\pi^0 e^+\) lifetime induced by \(X\)-boson exchange to satisfy the current experimental limit, \(\tau (p \to\pi^0 e^+) > 2.4\times 10^{34}~\mathrm{years}\)~\cite{Super-Kamiokande:2020wjk}. We take the Higgsino mass to be \(\mu=100~{\rm TeV}\), set \(\tan\beta=3\), and assume a common mass \(m_{\rm SUSY}=100~{\rm TeV}\) for the sfermions and the heavy Higgs bosons. The precise choice of \(m_{\rm SUSY}\) has only a negligible impact on the results. For each point in the \(M_2/M_1\)--\(M_3/M_2\) plane, the overall gaugino mass scale is fixed by requiring \(\delta m=400~{\rm keV}\). Again, the red and blue stars indicate the gaugino-mass ratios expected from anomaly mediation and gaugino-mass unification, respectively. As seen in Fig.~\ref{fig:lambdasigma_bound}, the present constraint on dimension-six proton decay can be avoided for \(\lambda_\Sigma\lesssim {\cal O}(10^{-2})\) throughout the parameter region of interest. This behavior follows directly from the relation between \(M_X\), \(M_{\rm GUT}\), and \(\lambda_\Sigma\): decreasing \(\lambda_\Sigma\) raises \(M_X\) at fixed \(M_{\rm GUT}\), thereby suppressing the gauge-boson-mediated decay rate, which scales as \(M_X^{-4}\). In the following analysis, we assume that \(\lambda_\Sigma\) is sufficiently small to satisfy this constraint. Under this assumption, the contribution of \(\mathrm{SU}(5)\) gauge-boson exchange to \(p\to K^+\bar{\nu}\) is also highly suppressed. We therefore neglect the dimension-six contribution to \(p\to K^+\bar{\nu}\) in the analysis below and focus on the color-triplet-Higgs-induced dimension-five contribution.

It is worth noting, however, that if \(\lambda_\Sigma\) lies close to the upper bound shown in Fig.~\ref{fig:lambdasigma_bound}, the \(X\)-boson-mediated \(p\to\pi^0 e^+\) rate can be close to the current experimental sensitivity. In such a case, this mode may become observable in future proton-decay searches at Hyper-Kamiokande~\cite{Hyper-Kamiokande:2018ofw}, providing a complementary test of the heavy-Higgsino scenario.

\begin{figure}
  \centering
  \begin{subfigure}[t]{0.48\textwidth}
    \centering
    \includegraphics[width=\linewidth]{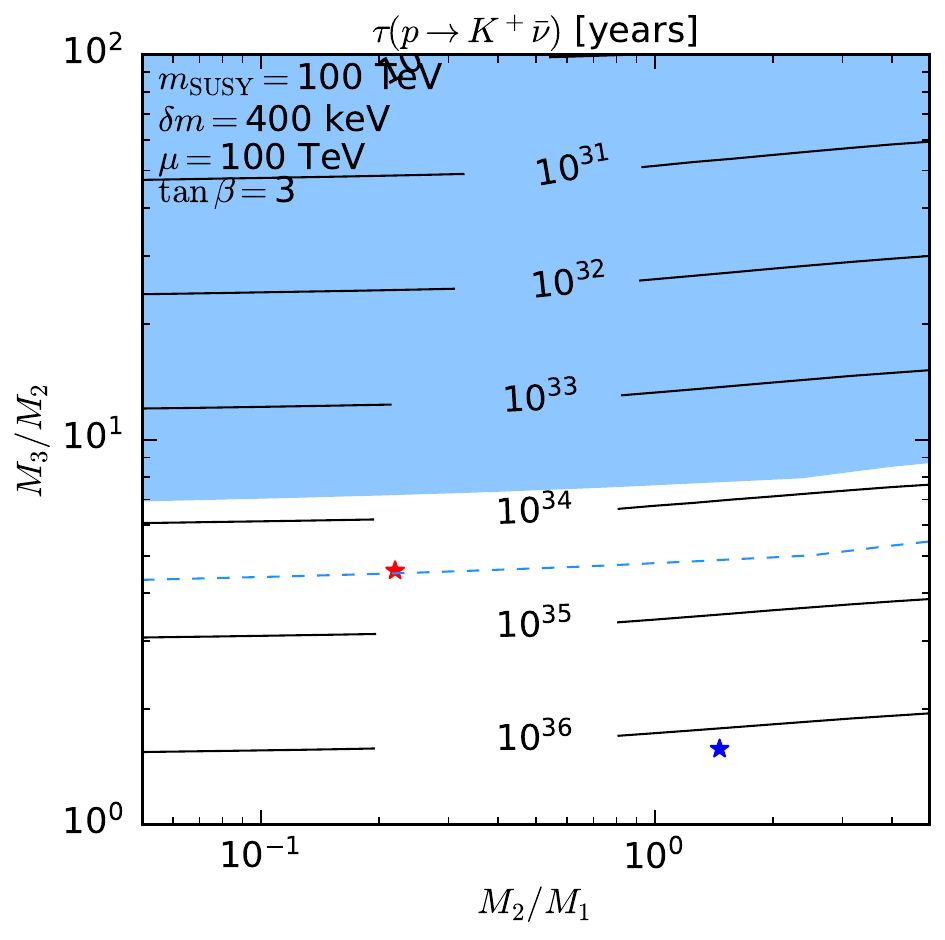}
    \caption{ \(m_{\rm SUSY}=100\)~TeV.
    }
    \label{fig:tau_Knu_100TeV}
  \end{subfigure}
  \hfill
  \begin{subfigure}[t]{0.48\textwidth}
    \centering
    \includegraphics[width=\linewidth]{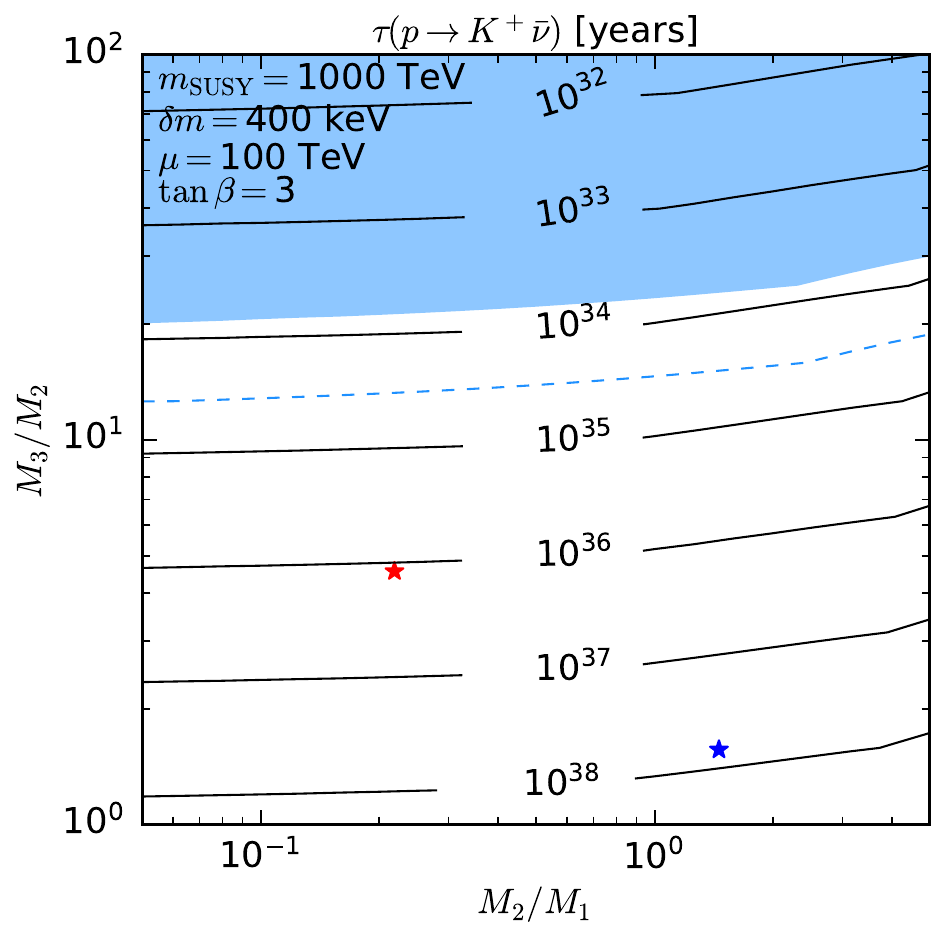}
    \caption{
    \(m_{\rm SUSY}=1000\)~TeV.
    }
    \label{fig:tau_Knu_1000TeV}
  \end{subfigure}
  \hfill
  \begin{subfigure}[t]{0.48\textwidth}
    \centering
    \includegraphics[width=\linewidth]{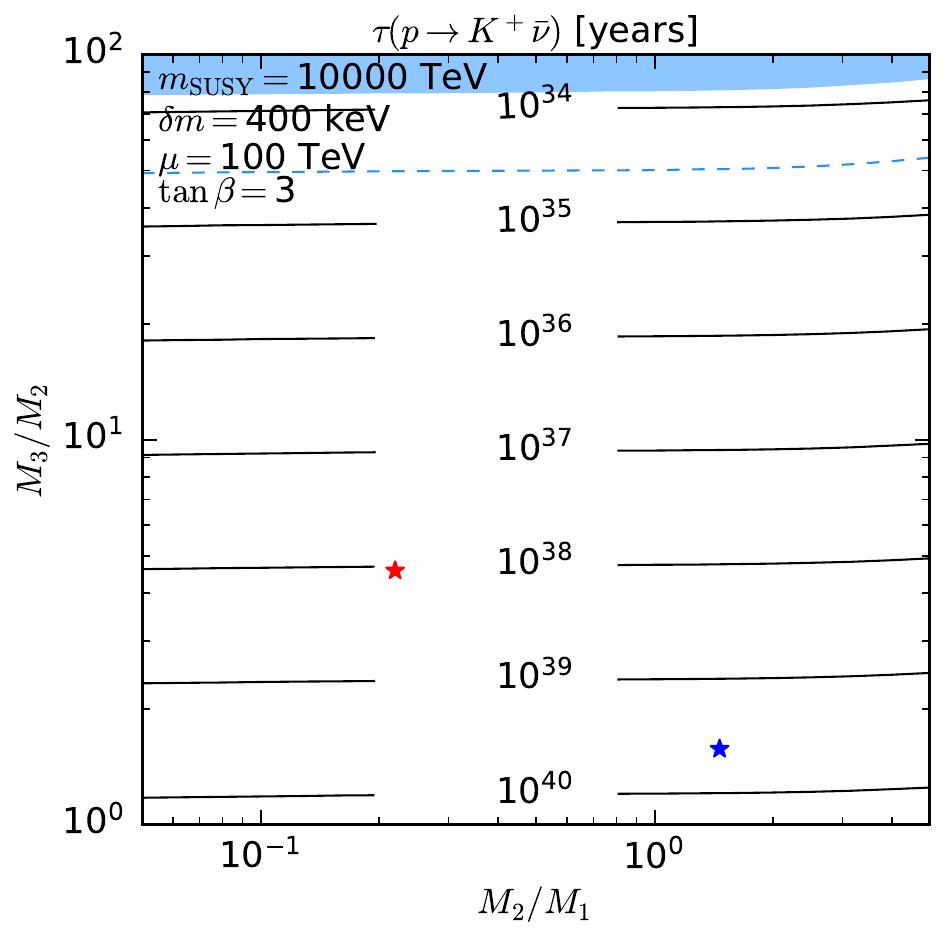}
    \caption{
    \(m_{\rm SUSY}=10000\)~TeV.
    }
    \label{fig:tau_Knu_10000TeV}
  \end{subfigure}
  \caption{
  Lifetime of the dimension-five proton decay mode \(p\to K^+\bar{\nu}\) in the \(M_2/M_1\)--\(M_3/M_2\) plane for (a) \(m_{\mathrm{SUSY}} = 100~\mathrm{TeV}\), (b) 1000~TeV, and (c) 10000~TeV. We take \(\mu=100~{\rm TeV}\), \(\tan\beta=3\), \(\varphi_i = 0\), and determine the overall gaugino mass scale by imposing  \(\delta m =400~{\rm keV}\). The blue-shaded region is excluded by the current experimental bound, \(\tau(p\to K^+\bar{\nu}) > 6.6\times 10^{33}\)~years~\cite{Super-Kamiokande:2014otb, Takhistov:2016eqm}, while the blue dashed line indicates the projected \(90\%\) CL sensitivity for a \(1.9~{\rm Megaton}\cdot{\rm year}\) exposure of Hyper-Kamiokande~\cite{Hyper-Kamiokande:2018ofw}. The red and blue stars indicate the gaugino-mass ratios expected from anomaly mediation~\cite{Randall:1998uk,Giudice:1998xp} and gaugino-mass unification, respectively. 
  }
  \label{fig:tau_Knu}
\end{figure}

We finally turn to the dimension-five proton decay, which provides one of the most important predictions of the present scenario. Unlike the dimension-six proton decay, the dimension-five proton-decay rate depends on the sfermion mass scale through the dressing diagrams. In this section, from a phenomenological point of view, we consider sfermion masses over the relatively broad range \(M_{\widetilde H}\sim 10^5~{\rm GeV} \lesssim m_{\widetilde f} \lesssim 10^7~{\rm GeV}\sim M_2\). This range may extend beyond the sfermion mass scales suggested by some specific underlying supersymmetry-breaking scenarios. We nevertheless adopt it here in order to explore the phenomenologically viable parameter space as broadly as possible and to assess the full range of possible implications for proton-decay experiments. The characteristic sfermion mass scales expected in concrete supersymmetric scenarios will be discussed in Sec.~\ref{eq:SUSYscenarios}.

Figure~\ref{fig:tau_Knu} shows the lifetime of
\(p\to K^+\bar{\nu}\) induced by the dimension-five operators in the \(M_2/M_1\)--\(M_3/M_2\) plane for (a) \(m_{\mathrm{SUSY}} = 100~\mathrm{TeV}\), (b) 1000~TeV, and (c) 10000~TeV. We set \(\mu=100~{\rm TeV}\), \(\tan\beta=3\), and take the GUT phases \(\varphi_i\) defined in Eq.~\eqref{eq:hij} to vanish, \(\varphi_i=0\). For each point, the overall gaugino mass scale is determined by imposing \(\delta m=400~{\rm keV}\). As discussed above, we assume that \(\lambda_\Sigma\) is sufficiently small that the contribution from \(\mathrm{SU}(5)\) gauge-boson exchange is negligible for this decay mode. The red and blue stars indicate the gaugino-mass ratios expected from anomaly mediation and gaugino-mass unification, respectively. These plots show that the predicted proton lifetime decreases significantly as \(M_3/M_2\) is increased. This behavior is mainly driven by the corresponding change in the color-triplet Higgs mass. As shown in Fig.~\ref{fig:gutmasses_mhc}, increasing \(M_3/M_2\) lowers \(M_{H_C}\), thereby enhancing the dimension-five Wilson coefficients, which are proportional to \(M_{H_C}^{-1}\). Consequently, the same hierarchy among the gaugino masses that improves gauge coupling unification also tends to enhance the dimension-five proton-decay rate. Remarkably, there is a sizable region of parameter space that is consistent with the present \(p\to K^+\bar{\nu}\) bound (blue-shaded region), \(\tau(p\to K^+\bar{\nu}) > 6.6\times 10^{33}\)~years~\cite{Super-Kamiokande:2014otb, Takhistov:2016eqm}, but lies within the projected sensitivity of Hyper-Kamiokande (blue dashed line), \(3.2\times 10^{34}~\mathrm{years}\)~\cite{Hyper-Kamiokande:2018ofw}. In particular, the gaugino-mass ratios suggested by anomaly mediation, indicated by the red star, lie within this experimentally interesting region for \(m_{\mathrm{SUSY}}=100~{\rm TeV}\). For heavier sfermion masses, a larger hierarchy between the gluino and wino masses is required to bring the proton lifetime within experimental reach: \(M_3/M_2\gtrsim 10\) for \(m_{\mathrm{SUSY}}=1000~{\rm TeV}\) and
\(M_3/M_2\gtrsim 50\) for \(m_{\mathrm{SUSY}}=10000~{\rm TeV}\). The heavy-Higgsino interpretation of the LZ event can therefore lead to a proton lifetime accessible to the next generation of proton-decay experiments.\footnote{The projected sensitivities to the \(p\to K^+\bar{\nu}\) lifetime are \(1.3\times 10^{34}~\mathrm{years}\) for DUNE~\cite{DUNE:2020fgq} and \(9.6\times 10^{33}~\mathrm{years}\) for JUNO~\cite{JUNO:2022qgr}.} This provides a direct connection between the hundred-keV-scale mass splitting relevant for dark-matter direct detection and baryon-number violation associated with physics near the GUT scale.

\section{Theoretical implications}
\label{sec:scenarios}

\subsection{Cosmology}

A Higgsino with a mass of \(\mathcal{O} (10^5)~{\rm GeV}\) cannot reproduce the observed dark-matter abundance in the standard thermal history.  In standard cosmology, a nearly pure thermal Higgsino accounts for the observed dark-matter density for a mass of \(M_{\widetilde H}\simeq 1.1~{\rm TeV}\)~\cite{Cirelli:2007xd}. A Higgsino as heavy as that considered in this work therefore gives a relic abundance far in excess of the observed value.

One simple solution to this problem was discussed in Ref.~\cite{Feldstein:2013uha}.\footnote{For other recent proposals, see Refs.~\cite{Langhoff:2026ujr, Pallis:2026jdc}.} During perturbative reheating after inflation, the temperature of the thermal bath can be considerably higher than the reheating temperature \(T_R\)~\cite{Giudice:2000ex}.  Denoting the maximum temperature during reheating by \(T_{\rm max}\), a dark-matter particle heavier than \(T_R\) can be produced out of equilibrium from scatterings in the thermal bath while never reaching chemical equilibrium.  Its abundance is then Boltzmann suppressed and can be much smaller than the conventional thermal relic abundance.  For a fermionic electroweak doublet with hypercharge \(Y=1/2\), as appropriate for the Higgsino at temperatures much larger than its neutral-state mass splitting, Ref.~\cite{Feldstein:2013uha} finds that the relic abundance is controlled approximately by
\begin{equation}
 x_{\rm eff}
 =
 -\frac{1}{2}
 \ln\left[
 \Gamma(9,2x_{\rm max})
 \right]
 +
 \frac{7}{2}\ln x_R\,,
 \label{eq:xeff}
\end{equation}
where \(\Gamma(s,x)\) denotes the upper incomplete gamma function,
\begin{equation}
  \Gamma(s,x)\equiv \int_x^\infty dt\,t^{s-1}e^{-t} \,,
\end{equation}
and 
\begin{equation}
 x_R\equiv \frac{M_{\widetilde H}}{T_R}\,,
 \qquad
 x_{\rm max}\equiv\frac{M_{\widetilde H}}{T_{\rm max}}\,.
\end{equation}
The first term on the right-hand side of Eq.~\eqref{eq:xeff} represents the Boltzmann suppression associated with the finite maximum temperature \(T_{\rm max}\), while the second term accounts for the dilution of the dark-matter abundance produced during reheating by the entropy production associated with the subsequent inflaton decay. The observed dark-matter density is reproduced for \(x_{\rm eff}\simeq 23\).  This condition corresponds approximately to
\begin{equation}
 30
 \lesssim
 \frac{M_{\widetilde H}}{T_R}
 \lesssim
 10^{3.5}.
 \label{eq:TRwindow}
\end{equation}
For the Higgsino mass of interest in the present work, \(M_{\widetilde H}\simeq10^5~{\rm GeV}\), this translates into
\begin{equation}
 30~{\rm GeV}
 \lesssim
 T_R
 \lesssim
 3\times10^3~{\rm GeV}.
 \label{eq:TRheavyhiggsino}
\end{equation}
Thus, the heavy Higgsino suggested by the LZ event can have the correct dark-matter abundance if the reheating temperature is substantially below the Higgsino mass.  For \(T_{\rm max}\gtrsim M_{\widetilde H}/4\simeq2.5\times10^4~{\rm GeV}\), the relic abundance becomes approximately insensitive to \(T_{\rm max}\), and the required reheating temperature approaches the lower end of Eq.~\eqref{eq:TRheavyhiggsino}, \(T_R\simeq {\cal O}(30)~{\rm GeV}\). In this regime, the final Higgsino abundance is therefore controlled mainly by the dilution due to entropy production during reheating, rather than by the Boltzmann suppression associated with \(T_{\rm max}\). For a lower maximum temperature, the additional Boltzmann suppression allows a larger \(T_R\), up to the TeV range. Even for such low reheating temperatures, the observed baryon asymmetry of the Universe may still be generated through non-thermal leptogenesis~\cite{Fukugita:1986hr} through the inflaton decay into right-handed neutrinos~\cite{Lazarides:1990huy,Kumekawa:1994gx,Asaka:1999yd,Asaka:1999jb}, whose subsequent CP-violating decays produce a lepton asymmetry. For discussions of leptogenesis at low reheating temperatures, see, e.g., Refs.~\cite{Zhang:2023oyo, Garcia:2026ulw}. We could also consider a low-scale leptogenesis scenario~\cite{Pilaftsis:1997jf, Pilaftsis:2003gt, Akhmedov:1998qx, Asaka:2005pn}. A more detailed construction of a concrete cosmological scenario realizing both the heavy-Higgsino dark-matter abundance and successful baryogenesis is left for future work.

\subsection{Supersymmetric scenarios}
\label{eq:SUSYscenarios}

The results obtained in Sec.~\ref{sec:results} may also provide a clue to the underlying mechanism of supersymmetry breaking.  We have found that, for a spectrum satisfying
\begin{equation}
 |\mu| \lesssim m_{\tilde{f}} < |M_i|,
 \label{eq:susy-hierarchy}
\end{equation}
with the gluino significantly heavier than the wino, the color-triplet Higgs mass can remain close to the conventional GUT scale, \(M_{H_C}\sim 10^{16}~{\rm GeV}\).  In this region, gauge coupling unification is considerably improved, while the dimension-five \(p\to K^+\bar{\nu}\) lifetime can fall within the reach of future proton-decay experiments.  It is therefore interesting to ask whether such an apparently peculiar supersymmetric spectrum can arise naturally from an underlying theory.

The required hierarchy among the gaugino masses is particularly suggestive. As is evident from Eq.~\eqref{eq:amsb-gaugino}, anomaly-mediated supersymmetry breaking~\cite{Giudice:1998xp,Randall:1998uk} naturally yields a gluino that is substantially heavier than the wino. Recall that the red stars shown in Figs.~\ref{fig:gutmasses}, \ref{fig:lambdasigma_bound}, and \ref{fig:tau_Knu} correspond to these anomaly-mediated gaugino-mass ratios. It is therefore particularly interesting that this simple and well-motivated pattern lies in the region where the dimension-five proton-decay rate becomes experimentally accessible.  There is, however, an additional issue if Eq.~\eqref{eq:amsb-gaugino} is responsible for the gaugino masses. Since anomaly-mediated gaugino masses are loop-suppressed relative to the gravitino mass, the \(M_2\sim 10^7~{\rm GeV}\) scale suggested by the Higgsino mass splitting corresponds parametrically to
\begin{equation}
  m_{3/2}
  \sim
  \frac{16\pi^2}{g_2^2}M_2
  \sim {\cal O}(10^9)~{\rm GeV}.
\end{equation}
Generic gravity-mediated scalar masses of order \(m_{3/2}\) would then be far above the sfermion masses considered in this work. A mechanism that protects the squarks and sleptons from acquiring \({\cal O}(m_{3/2})\) masses is therefore required, as we discuss below. More generally, non-universal gaugino mass spectra can arise in a variety of supersymmetry-breaking frameworks. Examples include models in which the gauge kinetic function depends on GUT-breaking Higgs fields~\cite{Ellis:1985jn, Drees:1985bx, Martin:2009ad}, mixed modulus--anomaly mediation~\cite{Choi:2005uz, Choi:2007ka}, and string-inspired constructions~\cite{Corsetti:2000yq}. Such scenarios may therefore provide alternative realizations of the gaugino-mass hierarchy relevant to the present analysis.

One possibility for suppressing the sfermion masses is that the gaugino masses constitute the dominant source of supersymmetry breaking in the MSSM sector, with the scalar soft masses generated radiatively from them. Well-known examples of this possibility are gaugino mediation~\cite{Kaplan:1999ac,Chacko:1999mi,Schmaltz:2000gy} and no-scale supergravity models~\cite{Cremmer:1983bf,Ellis:1983sf, Ellis:1984bm,Lahanas:1986uc,Ellis:2017djk}. In such frameworks, the sfermion masses are loop suppressed relative to the gaugino masses. It is, however, not straightforward to obtain a large hierarchy between the squark masses and the electroweak gaugino masses, since the squark masses receive sizable radiative contributions from the gluino and are therefore enhanced when the gluino is heavy.

An attractive alternative is that the MSSM matter chiral superfields are realized as Nambu--Goldstone multiplets. The idea that quarks and leptons may arise as quasi-Nambu--Goldstone fermions was proposed in Ref.~\cite{Buchmuller:1982tf} and developed further in Ref.~\cite{Kugo:1983ai}.\footnote{For a recent related work, see Ref.~\cite{Yanagida:2016kag}. } In particular, Ref.~\cite{Kugo:1983ai} considered a supersymmetric nonlinear sigma model based on the K\"ahler coset \(\mathrm{E}_7/\mathrm{SU}(5)\times\mathrm{SU}(3)\times\mathrm{U}(1)\). Remarkably, the Nambu--Goldstone chiral multiplets associated with this coset contain three copies of \(\mathbf{10}\oplus\overline{\mathbf{5}}\) of \(\mathrm{SU}(5)\), corresponding precisely to the three observed generations of quarks and leptons, together with a \(\mathbf{5}\).\footnote{An additional \(\overline{\mathbf{5}}\) must be introduced to remove the nonlinear sigma-model anomaly~\cite{Yanagida:1985jc}. This multiplet may form a pair of Higgs multiplets with the \(\mathbf{5}\) mentioned in the text.} The \(\mathrm{SU}(3)\) factor organizes the three families, so that the generation structure itself is related to the geometry of the coset space rather than being imposed by hand. If the MSSM matter multiplets have such an origin, the same symmetry structure that accounts for the three-generation structure can simultaneously protect their scalar partners from large soft masses. In such a construction, the squarks and sleptons are associated with
Nambu--Goldstone modes, and their soft masses can be protected from generic tree-level supergravity contributions. Their masses are then generated by effects that explicitly break the underlying global symmetry. In general, there may be several sources of such explicit global-symmetry breaking beyond the SM gauge and Yukawa interactions, and both the overall sfermion mass scale and the detailed spectrum therefore depend on the structure and size of these global-symmetry breaking effects. As a simple measure of the explicit global-symmetry breaking, we introduce the ratio \(\epsilon_G \equiv {m_{\rm SUSY}}/{m_{3/2}}\), where \(m_{3/2}\simeq 10^9~{\rm GeV}\) in the anomaly-mediated scenario considered above. In this work, we consider \(100~{\rm TeV}\lesssim m_{\rm SUSY}\lesssim 10000~{\rm TeV}\), corresponding to \(\epsilon_G\sim 10^{-4}\)--\(10^{-2}\). The lower end of this range is motivated by the
requirement that the lightest sfermion be heavier than the Higgsino, so that the \(\sim 100~{\rm TeV}\) Higgsino can remain the stable LSP and constitute dark matter. At the same time, taking \(m_{\rm SUSY}< 10000~{\rm TeV}\), well below the gaugino masses of order \(10^7~{\rm GeV}\), allows the gauginos to decay efficiently into sfermions and SM fermions. These requirements are therefore consistent with a picture in which the sfermion masses arise from small explicit breaking of the quasi-Nambu--Goldstone symmetry. This also makes proton decay particularly valuable as a probe of the underlying dynamics: since the dimension-five proton-decay rate depends sensitively on the sfermion spectrum, future proton-decay measurements could provide information on the pattern and magnitude of the global-symmetry breaking responsible for generating the sfermion masses.

Finally, the dark-matter interpretation imposes an additional nontrivial requirement on this picture. Since the sfermions are assumed to be lighter than the gauginos, the latter can decay into quark--squark or lepton--slepton final states and therefore cannot serve as dark matter. This in turn requires the Higgsino to be the lightest supersymmetric particle (LSP). Taken together, the spectrum suggested by the LZ event may therefore be pointing not only to a particular pattern of supersymmetry breaking, but also to more fundamental physics governing the origin and hierarchy of the supersymmetric mass parameters.

\section{Conclusion}
\label{sec:conclusion}

In this paper, we have studied the implications of the heavy-Higgsino interpretation of the LZ high-recoil event for supersymmetric grand unification. This interpretation favors a Higgsino mass of \(\mathcal{O}(10^5)~{\rm GeV}\) together with a neutral-Higgsino mass splitting of a few hundred keV, which in turn points to electroweak gaugino masses of \(\mathcal{O} (10^7)~{\rm GeV}\). Although such a heavy supersymmetric spectrum might naively appear unfavorable for supersymmetric GUTs, we have shown that this expectation need not hold even in minimal supersymmetric \(\mathrm{SU}(5)\).

By including the supersymmetric and GUT-scale threshold corrections to the gauge couplings, we determined the color-triplet Higgs mass \(M_{H_C}\) and the combination \(M_{\rm GUT}=(M_X^2M_\Sigma)^{1/3}\).  We found that the large Higgsino mass tends to raise \(M_{H_C}\), while a hierarchy with the gluino significantly heavier than the wino lowers it.  As a result, \(M_{H_C}\) can remain close to the conventional GUT scale, \(M_{H_C}\sim10^{16}~{\rm GeV}\), even for the very heavy gaugino spectrum suggested by the LZ event.  In the same region, \(M_{H_C}\) and \(M_{\rm GUT}\) become close to each other, and the three gauge couplings exhibit a high degree of unification.  

We have also examined proton decay in this setup.  Since gauge coupling unification fixes only the combination \(M_{\rm GUT}\), the \(\mathrm{SU}(5)\) gauge-boson mass \(M_X\) can be raised by taking the adjoint-Higgs self-coupling \(\lambda_\Sigma\) sufficiently small. For \(\lambda_\Sigma\lesssim{\cal O}(10^{-2})\), the current bound on the dimension-six mode \(p\to\pi^0e^+\) can readily be satisfied. On the other hand, the dimension-five mode \(p\to K^+\bar{\nu}\) provides a particularly interesting prediction. When the gluino is significantly heavier than the wino, the resulting reduction of \(M_{H_C}\) enhances the dimension-five proton-decay rate, and a sizable parameter region consistent with the present bound lies within the projected sensitivities of Hyper-Kamiokande, JUNO, and DUNE. Thus, the same gaugino-mass hierarchy that predicts \(M_{H_C}\sim 10^{16}~{\rm GeV}\) can simultaneously make proton decay observable in next-generation experiments.

Finally, the preferred supersymmetric spectrum may provide important clues to the underlying mechanism of supersymmetry breaking. A hierarchy \(m_{\widetilde f}<|M_i|\) can arise in frameworks such as gaugino mediation or no-scale supergravity, although obtaining a sizable separation between the squark and electroweak-gaugino masses is nontrivial when the gluino is heavy. The non-universal gaugino spectrum required in our scenario, in particular \(M_3\gg M_2\), can naturally arise in anomaly mediation as well as in other mechanisms generating non-universal gaugino masses. An especially attractive possibility is that the MSSM matter fields are realized as Nambu--Goldstone multiplets, in which case their scalar soft masses can be protected from generic supergravity contributions and are instead controlled by explicit breaking of the underlying global symmetry. The resulting sfermion spectrum can then be probed through dimension-five proton decay. The dark-matter interpretation further requires the Higgsino to lie below the sfermions and become the LSP, leading to the characteristic hierarchy \(|\mu|\lesssim m_{\widetilde f}<|M_i|\). Taken together, the spectrum suggested by the LZ event may therefore provide information not only on supersymmetry breaking itself, but also on the more fundamental dynamics governing the hierarchy of supersymmetric mass parameters.

\section*{Acknowledgments}

The work of S.H. was supported in part by Prof. Jason L. Evans and Prof. Yuichiro Nakai, funded by the National Natural Science Foundation of China.
The work of N.N. was supported in part by the Grant-in-Aid for Scientific Research C (No. 25K07314). T.~T.~Y. is supported by MEXT Grant No.~24H02244 and the World Premier International Research Center Initiative (WPI), MEXT, Japan (Kavli IPMU).

\newpage
\bibliographystyle{utphysmod}
\bibliography{ref}


\end{document}